\documentclass[preprint,12pt,3p]{elsarticle}

\usepackage{amssymb}
\usepackage{graphicx}
\usepackage{subfig}
\usepackage{amsmath}
\usepackage{algorithm}
\usepackage{algorithmic}
\usepackage{float}
\usepackage{multirow}
\usepackage{scrextend}
\usepackage{tablefootnote}
\usepackage{adjustbox}
\usepackage{color, colortbl}
\definecolor{Gray}{gray}{0.9}

\renewcommand{\phi}{\varphi}
\newcommand{\argmin}{\arg\!\min}
\journal{Knowledge-Based Systems}

\begin{document}

\begin{frontmatter}

\title{Feature Assisted bi-directional LSTM Model for Protein-Protein Interaction Identification from Biomedical Texts}
\author{Shweta Yadav, Ankit Kumar, Asif Ekbal, Sriparna Saha and Pushpak Bhattacharyya}
\cortext[CorresAuthor]{Asif Ekbal}
\ead{asif.ekbal@gmail.com,asif@iitp.ac.in}
\address{Department of Computer Science and Engineering \\Indian Institute of Technology Patna (IIT Patna)\\Patna, India}

\begin{abstract}
Knowledge about protein-protein interactions is essential in understanding the biological processes such as metabolic pathways, DNA replication, and transcription etc. However, a majority of the existing Protein-Protein Interaction (PPI) systems are dependent primarily on the scientific literature, which is yet not accessible as a structured database. Thus, efficient information extraction systems are required for identifying PPI information from the large collection of biomedical texts. Most of the existing  systems model the PPI extraction task as a classification problem and are tailored to the handcrafted feature set including domain dependent features. In this paper, we present a novel method based on deep bi-directional long short-term memory (B-LSTM) technique that exploits word sequences and dependency path related information to identify PPI information from text. This model leverages joint modeling of proteins and relations in a single unified framework, which we name as Shortest Dependency Path B-LSTM (sdpLSTM) model.
We perform experiments on two popular benchmark PPI datasets, namely AiMed \& BioInfer. The evaluation shows the F1-score values of $86.45\%$ and $77.35\%$ on AiMed and BioInfer, respectively. Comparisons with the existing systems show that our proposed approach attains state-of-the-art performance.
\end{abstract}


\end{frontmatter}


\section{Introduction}
\label{intro}
The study of the Protein-Protein Interaction (PPI) is crucial in understanding the biological process, such as DNA replication and transcription, metabolic pathways and cellular organization. 
Owing to this fact, several databases have been manually curated to cache protein interaction information such as MINT \cite{zanzoni2002mint}, BIND \cite{bader2003bind}, and SwissProt \cite{bairoch2000swiss} in structured and standard formats. However, the rapid growth of biomedical literature has shown a significant gap between the availability of protein interaction article and its automatic curation. As such, a majority of the protein interaction information is still uncovered in the textual contents of biomedical literature. Moreover, the growth in biomedical literature is at an exponential pace. In the last 20 years, the overall size of MEDLINE has increased at a ~4.2\% compounded annual growth rate. There is ~3.1\% compounded annual growth rate in the number of new entries in MEDLINE database. MEDLINE currently has more than $6,000,000$ publications, which are more than three millions than those  published in the last 5 years alone \cite{hunter2006biomedical}. Hence, owing to the exponential rise \cite{lu2011pubmed,khare2014accessing} and complexity of the biological information, the necessity for intelligent information extraction techniques to assist biologist in detecting, curating and maintaining database is becoming crucial. This has lead to a surge in the interest of Biomedical Natural Language Processing (BioNLP) community for automatic detection and extraction of PPI information. 

Determining PPIs in the scientific text is the process of recognizing how two or more proteins in the given biomedical sentence are related. We exemplify interaction types between protein pairs in Table-\ref{example} where protein (\textit{Bnrlp-Rho4p}) forms the interacting protein pair and the (\textit{Bnrlp-Rho1p}) is non-interacted protein pairs. \\
\begin{table}[h!]
\centering
\resizebox{\textwidth}{!}{%
\begin{tabular}{lccc}
\hline
Sentence                                                                  & Protein Entities      & Interacted Protein Pair & Non-interacted Protein Pair \\ \hline
\begin{tabular}[c]{@{}l@{}}Bnrlp interacts with another Rho family member, Rho4p,\\  but not with Rho1p.\end{tabular} & Bnrlp, Rho4p, Rho1p & Bnrlp-Rho4p             & Bnrlp-Rho1p                 \\ \hline
\begin{tabular}[c]{@{}l@{}}These data demonstrate that Stat3 but not Stat1 or Stat5 is\\ directly recruited to the ligand-activated IL-10 receptor.\end{tabular} & Stat3, Stat1, Stat5, IL-10 & Stat3-IL-10         & \begin{tabular}[c]{@{}l@{}}Stat3-Stat1, Stat3-Stat5, Stat3-IL-10,\\  Stat1-Stat5, Stat1-IL-10, Stat5-IL-10\end{tabular}                 \\ \hline
\end{tabular}%
}
\caption{Exemplar description of protein protein interaction in a sentence}
\label{example}
\end{table}

Majority of the existing systems look upon this task as a binary classification problem by identifying whether any interaction occurs between a pair of proteins or not. One of the most explored techniques for PPI task includes kernel-based method \cite{bunescu2005shortest,airola2008graph}. The potentiality of the kernel-based method is due to the virtue of a large amount of carefully crafted features. However, extraction of these features relies on the other NLP tools such as ABNER \cite{settles2005abner}, MedT-NER \cite{saetre2007akane} or PowerBioNE \cite{zhou2004recognizing} and machine learning (ML) tool (SVM-light with Tree-Kernels). Recently, with the widespread usages of neural network based techniques in clinical and biomedical domain \cite{yadav2017entity,yadav2018multi,yadav2016deep,ekbal2016deep,kumar2016recurrent}, methods exploring latent features have emerged as strong alternative choices over the traditional machine learning based techniques. Some of the distinguished studies \cite{hua2016shortest,choi2016extraction} for PPI extraction tasks utilize convolution neural networks (CNNs) which have shown significant performance improvements over the existing state-of-art techniques. Some other popular neural network based models for relation extraction have been reported in \cite{miwa2016end,liu2015dependency}. But these systems are mostly applicable in identifying different relationships from newswire articles. Thus these approaches fail to produce a comparable performance on biomedical literature owing to the complexity of the biomedical text. Biomedical named entities do not have standard nomenclature, and this arbitrariness increases the difficulty in capturing the semantic relationships between the entities (proteins). Moreover, the different protein entities  often have similar names making it more difficult to capture the contextual information. 

Motivated by these observations, in this paper we propose a Shortest Dependency Path based Bi-directional Long Short Term Memory (sdpLSTM) architecture \cite{rnn1} to identify PPI pairs from the text. The proposed method differs from the previous studies in two facets: First, utilizing the dependency relationships between the protein names, we generate the Shortest Dependency Path (SDP) of the sentences. This facilitates us to create more syntax-aware inferences about the capabilities of the proteins in a sentence in comparison to the technique developed based on classical kernel-based method. Second, we investigate the significance of Part-of-Speech (PoS) and position embedding features in improved learning of the sdpLSTM. 
In contrast to the systems proposed by \cite{hua2016shortest} and \cite{choi2016extraction}, we employ LSTM based neural network models \cite{rnn1} instead of Convolutional Neural Network (CNN) \cite{lecun1995convolutional}. In CNN, features are generated by performing pooling over the entire sentence based on continuous $n$ grams, where $n$ refers to the filter size. This puts 
constraints on longer sentences where long-term dependencies exist. Our method circumvents the shortcoming of CNN architecture by utilizing pooling techniques for encoding variable length features.
In general, Bi-LSTM can keep track of preceding and succeeding words. As such, when pooling operation is performed on the output of sdpLSTM, we obtain optimal features from the entire sentence carrying the information not just on $n$-grams but the complete context of the sentence.

In contrary, the existing methods \cite{miwa2016end, peng2017deep} generally consider a whole sentence as the input. The drawback of these existing techniques is that such representations fail to describe the relationships of two
target entities which appear in the same sentence at a far distance (i.e. long distant). 
Considering these problems, in our proposed technique we exploited dependency parsing related feature to examine the sentence and capture the Shortest Dependency Path to generate SDP based word embedding. In order to further inject the explicit linguistic information and boost the performance of the LSTM architecture, we have included the PoS information of SDP based words to assist the LSTM based network. The position w.r.t protein and part-of-speech (PoS) are prominent features in identifying the protein interaction information. PoS provides useful evidence that helps to detect important grammatical properties. Words assigned with same PoS posses similar syntactic behavior which provides an important clue to the system for inferencing the interaction between the protein pair. \\
The basic structure of a sentence can be obtained by determining the position of protein-word and the word occurring in its vicinity which provides pivotal clues to identify interactions in sentences. The extraction of SDP based word embeddings rather than full sentence embedding and its usage as an input to LSTM network in an amalgamation with the position and PoS feature is the core contribution of our proposed work.

The key attributes of the proposed work are summarized as below: 
\begin{enumerate}
\itemsep0.5em
\item We propose the shortest dependency path based Bi-LSTM model (sdpLSTM) that provides state-of-the-art performance for relation extraction.
\item We explore latent features like Part-of-Speech (PoS) and position w.r.t proteins which were found to be effective in extracting protein-protein pairs.
\item We demonstrate that word embedding models learned on the PubMed, PMC and Wikipedia corpus is more powerful than the internal embedding model or the model trained on general corpus such as the news corpus\footnote{https://code.google.com/archive/p/word2vec/}.
\item Evaluation on two different benchmark corpora, namely AiMed \& BioInfer establishes the fact that our proposed approach is generic in nature. Please note that these two datasets were created by following different protein annotation guidelines. 
\end{enumerate}
\section{Related Works}\label{relatedWork}
\begin{enumerate}
\item \textbf {Pattern-based Model:} Preliminary studies conducted by \cite{blaschke1999automatic} and \cite{ono2001automated} explored pre-specified patterns and rules for the PPI task. However, the system lacks in identifying complex cases such as complex relationships expressed in various coordinating and relational clauses. For sentences containing complex relations between three
or more entities, the approach usually yields erroneous results. For example, \\
``\textit{The gap1 mutant blocked stable association of
Ste4p with the plasma membrane, and the ste18 mutant
blocked stable association of Ste4p with both plasma membranes and internal membranes.}'' \\
In \cite{huang2004discovering} authors proposed a technique based on dynamic programming to automatically discover patterns. The system proposed in \cite{bunescu2005comparative} also studied the performance of rule-based algorithms. They developed two models, first one made use of rapier rule-based system and the other one relied on longest common subsequences.\\ 
\item \textbf {Using Dependency Parsing:} Here we describe the works that take into account more syntax aware approach such as full and partial (shallow) parsing. In the partial parsing, sentence structure is divided partially and dependencies are generated locally within the phrase. While in full parsing, the whole sentence is considered to capture dependencies, \cite{giuliano2006exploiting} developed the system solely based on the shallow syntactic information. They further incorporated kernel functions to combine information from the entire sentence and the local contexts around the interacting entities. The work reported in \cite{erkan2007semi} focused on extracting the SDP between the protein pairs by defining the cosine similarities and edit distance function via semi-supervised learning. Some of the other prominent works include the studies conducted by \cite{miyao2009evaluating} and \cite{Garg}. Other popular studies based on full parsing include the works as reported in \cite{temkin2003extraction,daraselia2004extracting,yakushiji2005biomedical}.\\
\item \textbf {Kernel-based Model:}
Bunescu and Mooney \cite{bunescu2005shortest} first proposed the idea of using kernel methods to extract PPI based on the SDP. Some of the effective kernel-based techniques for PPI task include graph kernel \cite{airola2008all}, bag-of-word (BoW) kernel \cite{saetre2007syntactic}, edit-distance kernel \cite{erkan2007semi}, all-path kernel \cite{airola2008graph} and tree kernel \cite{Ma,Zhang}.\\
\item \textbf {Deep Learning based Model:} Recent studies show the applicability of deep learning models for the PPI 
task \cite{hua2016shortest,choi2016extraction}. The work reported in \cite{hua2016shortest} made use of Convolutional Neural Network (CNN) for developing the PPI based system. 
\cite{choi2016extraction} proposed a CNN based model utilizing several handcrafted features exploiting lexical, syntactic and semantic level information in combination with word embeddings. 
\end{enumerate}
\begin{figure}[t]
\centering
\includegraphics[ height=6cm,width=\linewidth]{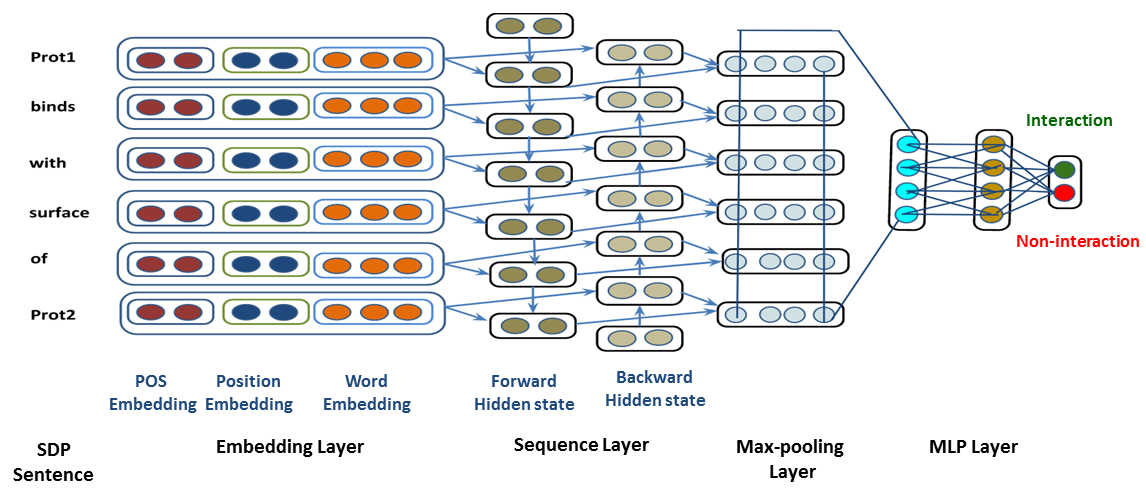}
\label{blstm}
\caption{Proposed model architecture for protein protein interaction. The input is the Shortest Dependency Path (SDP) between a pair of protein. The output of the model is the probability distribution over two class:`\textit{interaction}' and `\textit{non-interaction}'. (all the neurons representation are hypothetical)}
\end{figure}
\section{Method}\label{architecture}
In this study, we present a novel model to predict protein interaction pairs from the text. Our model leverages joint modeling of proteins and relations in a single model by exploiting Bi-directional Long Short Term Memory (Bi-LSTM) technique and propose a Shortest Dependency Path Bi-LSTM (sdpLSTM) model. 
Dependency between entities captures the information relevant for identifying the relations.  
Further, this architecture utilizes positional information of proteins in the sentence and the PoS embedding as the latent features in improved learning of sdpLSTM model. We begin by extracting SDP sentences and exploiting latent features. Embeddings are generated corresponding to each feature which is passed as input to the Bi-LSTM unit. We describe each phase in succeeding subsections.

\subsection{Shortest Dependency Path (SDP) }\label{dependency}
\begin{figure*}[t]
\centering
\includegraphics[width=\textwidth, height=3cm]{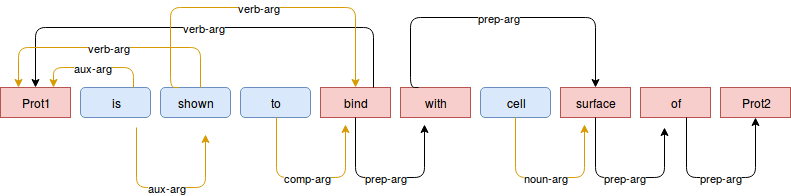}
\caption{The predicate argument of the example sentence ``\textit{Prot1 is shown to bind with cell surface of Prot2}." Here, the words represent the nodes and predicate argument relation is represented by edges. The red nodes form the SDP for the given sentence with the black arrow denoting the path to reach from `$Prot_1$' to `$Prot_2$'. The other words are represented in blue round-rectangular boxes that are not part of SDP. Thereby, the SDP for given sentence is ``\textit{Prot1 bind with surface of Prot2}'' }
\label{figure-1}
\end{figure*}
The input to the sdpLSTM is the SDP between a protein pair. For this purpose, we exploit the dependency parse tree of the sentence. It describes the syntactic constituent structure of the sentence by annotating edges with dependency types, e.g. \textit{subject}, \textit{auxiliary}, \textit{modifier} and captures the semantic \textit{predicate-argument} relationships between the words. In general, \cite{bunescu2005shortest} first proposed the idea of using dependency parse tree for relation extraction. They designed a kernel function exploring the shortest path between the entities to capture the relations. The main intuition behind this is based on the observation that shortest path reveals non-local dependencies within sentences which can help in capturing the relevant information from the sentence. The shortest path between the protein pair generally captures the essential information (aspects of sentence construction such as mood, modality and sometimes negation, which can significantly alter or even reverse the meaning of the sentence)
to identify their relationship. The approach proposed in \cite{culotta2004dependency} was proved to be significantly better over the dependency tree kernel-based model. We follow this idea to use SDPs for extracting protein interacting pairs.\\
As illustrated in Figure \ref{figure-1}, the word `bind' in SDP carries important information to predict the interaction between the protein pair. The dependency relation bounded here is by verb argument and as interaction verb carries essential evidence in PPI. For PPI task, capturing these dependency relations is important.\\
\indent For the purpose of extracting dependency relations, we use Enju Parser\footnote{http://www.nactem.ac.uk/enju/} which is a syntactic parser for English and can effectively analyze syntactic and/or semantic structures of biomedical text and provide with predicate-argument information.
We have generated a graph for every sentence that contains at least two protein entities where each word corresponds to the node of the graph and the edges between the nodes (dependency relation) are obtained by the parser. We utilize Breadth First Search (BFS) algorithm \cite{lee1961algorithm} to calculate the shortest distance between the protein pair. The words occurring between the SDP only takes part in the training instead of the whole words present in the sentences to generate SDP embedding.

\begin{table}[h!]
\centering
\resizebox{\textwidth}{!}{%
\begin{tabular}{cccccccccc}
\hline
\textbf{SDP Words} & \textbf{PoS} & \textbf{PoS Feature} & \begin{tabular}[c]{@{}c@{}}\textbf{PoS Feature} \\ \textbf{Encoding} \end{tabular} & \begin{tabular}[c]{@{}c@{}} \textbf{Relative position}\\ \textbf{from Prot$_1$} \end{tabular} & \begin{tabular}[c]{@{}c@{}} \textbf{Relative position}\\ \textbf{from Prot$_2$} \end{tabular} & \begin{tabular}[c]{@{}c@{}}  \textbf{Position}\\ \textbf{ Feature-1} \end{tabular} & \begin{tabular}[c]{@{}c@{}}  \textbf{Position}\\ \textbf{ Feature-2} \end{tabular} & \begin{tabular}[c]{@{}c@{}}  \textbf{Position Feature}\\ \textbf{ Encoding-1} \end{tabular} & \begin{tabular}[c]{@{}c@{}}  \textbf{Position Feature}\\ \textbf{ Encoding-2} \end{tabular} \\ \hline
Prot1 & NN & 10000000 & [0.00171600 \ldots0.0033500] & 0 & -6 & 0000000000 & 0000111111 & [0.03141600 \ldots0.9035500] & [0.1117600 \ldots0.0223500] \\ 
regulator & NN & 00000000 & [0.99121600 \ldots0.0233500] & 1 & -5 & 0000000001 & 0000011111 & [0.77171600 \ldots0.4858500] & [0.83191600 \ldots-0.1133500] \\ 
between & IN & 00100000 & [0.25191600 \ldots0.1739500] & 2 & -4 & 0000000011 & 0000001111 & [0.33171600 \ldots-0.8833500] & [0.58961600 \ldots0.7189200] \\
Interaction & NN & 10000000 & [0.17171219 \ldots0.7583350] & 3 & -3 & 0000000111 & 0000000111 & [0.75171600 \ldots0.5533500] &[0.99171600 \ldots0.7633500] \\ 
and & CC & 00100000 & [0.17001600 \ldots0.3030350] & 4 & -2 & 0000001111 & 0000000011 & [0.78117600 \ldots-0.033500] & [0.72171600 \ldots0.1233500] \\ 
repression & NN & 10000000 & [0.17858500 \ldots0.8835300] & 5 & -1 & 0000011111 & 0000000001 & [0.45897600 \ldots-0.0522500] & [0.7800100 \ldots0.3311500] \\ 
Prot2 & NN & 10000000 & [0.98581600 \ldots0.0263500] & 6 &  0 & 0000111111 & 0000000000 & [0.77451600 \ldots0.8985500] & [0.1745100 \ldots0.3323500] \\ \hline 
\end{tabular}%
}
\caption{Feature Encoding for sentence \textit{``Interaction between cell cycle regulator, Prot1, and Prot2 mediates repression of HIV-1 gene transcription.''}. Here, the words occurring in the vicinity of SDP are used to generate features.}
\label{feature}
\end{table}

\begin{table}[h!]
\centering
\resizebox{0.5\textwidth}{!}{%
\begin{tabular}{cccccccccccccccc}
\hline
 \textbf{Distance} &  \textbf{0} & \textbf{1} & \textbf{2} & \textbf{3} & \textbf{4} & \textbf{5} & \textbf{6} & \textbf{7} & \textbf{8} & \textbf{9} & \textbf{10} & \textbf{11-$\infty$} \\\hline
\multirow{10}{*}{\rotatebox{90}{\textbf{Binary representation}}}&0 & 0 & 0 & 0 & 0 & 0 & 0 & 0 & 0 & 0 & 1 & 1 \\
&0 & 0 & 0 & 0 & 0 & 0 & 0 & 0 & 0 & 1 & 1 & 1 \\
&0 & 0 & 0 & 0 & 0 & 0 & 0 & 0 & 1 & 1 & 1 & 1 \\
&0 & 0 & 0 & 0 & 0 & 0 & 0& 1& 1 & 1 & 1 & 1 \\
&0 & 0 & 0 & 0 & 0 & 0 & 1 & 1 & 1 & 1 & 1 & 1 \\
&0 & 0 & 0 & 0 & 0& 1 & 1 & 1 & 1 & 1 & 1 & 1 \\
&0 & 0 & 0 & 0 & 1 & 1 & 1 & 1 & 1 & 1 & 1 & 1 \\
&0 & 0 & 0 & 1 & 1 & 1 & 1 & 1 & 1 & 1 & 1 & 1 \\
&0 & 0 & 1 & 1 & 1 & 1 & 1 & 1 & 1 & 1 & 1 & 1 \\
&0 & 1 & 1 & 1 & 1 & 1 & 1 & 1 & 1 & 1 & 1 & 1 \\\hline
\end{tabular}
}
\caption{Binary representation of relative distance w.r.t the protein mention.}
\label{Table-3}
\end{table}
\subsection{Latent Feature Encoding Layer}\label{latent-feature}
Along with the SDP embedding,
we design domain-independent features to assist our model in becoming more generic and adaptable. We explore PoS and position of each word as a feature. An exemplar illustration of latent feature encoding is provided in Table-\ref{feature}. 
\begin{enumerate}
\item \textbf{PoS Feature:} This represents the PoS for each word occurring in the vicinity of SDP. We use 
Genia Tagger\footnote{http://www.nactem.ac.uk/GENIA/tagger/} to extract PoS information of each token. Every PoS tag is encoded as a unique eight dimension one hot vector which is fed to a neural network (NN) based encoder. Auto-encoder \cite{vincent2010stacked} is employed to transfer the sparse PoS features to the dense real-valued feature vectors. This converts one-hot representation to dense feature representation of dimension $8$. We use Adadelta optimizer \cite{adadelta} with loss function as a squared error to train our auto-encoder model. \\
Let $P$ represents the one hot vector of a PoS tag corresponding to each word. 
The auto-encoder learns the transition functions $\varphi$ and $\Omega$ such that reconstruction errors (squared errors) are minimized. The function $\varphi$ and $\Omega$ are called the encoder and the decoder function, respectively. Mathematically, it can be written as follows:
\begin{equation}
\varphi, \Omega=\argmin_{\varphi, \Omega}||P-P'||^2 
\end{equation}
where $\varphi:P \rightarrow Z$, $\Omega:Z \rightarrow P$.
\item \textbf{Position Feature:} This feature helps us in identifying the significant interacting tokens between the two target protein entities. 
The position feature computes the relative distances of a word with respect to the protein mentions. We use the binary representation of the distance. We extract this feature on SDP of the target protein pairs. It is a two-dimensional tuple denoting distances of these tokens from the two target proteins. 
For e.g., consider the following sentence: `Prot1 regulator between interaction and repression Prot2', the relative distances of the word `interaction' with respect to Prot1 and Prot2 are $-3$ and $3$, respectively. 
Relative distances are then mapped to 10-dimensional binary vectors. From Table-\ref{Table-3}, we can observe that more attention is given to the words near to the protein mentions, particularly to the words occurring in the vicinity of 10 surrounding words. Moreover, words whose relative distances exceed 10 are all treated equally.\\
Intuitively, the words which are nearer to the target words are more informative than the farther words. We perform experiments to determine the optimal dimension by varying the distance (from 5 to 12) of more informative words with respect to proteins as shown in Table \ref{context-window}. We notice that the system performs well when the maximum relative distance of the informative word is within the range of $10$ w.r.t the protein term. 
As we follow the binary representation of distance, therefore, the position feature is represented using a feature vector of $10$ dimensions.\\
Similar to PoS feature, every position feature is encoded as a 10-dimensional vector which is fed into an auto-encoder. Using the learned auto-encoder model, we convert the sparse position feature vector to a dense real valued feature vector of dimension $10$.
\end{enumerate}

\subsection{Embedding Layer}\label{embedding-layer}
Word embedding persuades a real-valued latent semantic or syntactic vector for each word from a large unlabeled corpus by using continuous space language models \cite{tang2014evaluating}. In embedding layer we obtain real-valued vector corresponding to each word of the SDP. Let us assume that we have a SDP sentence $S_{sdp}=\{w_1,w_2 \ldots w_N\}$ having size $N$, a pre-trained word embedding matrix $\textbf{M} \in \mathbb{R}^{d\times|V|} $. A real-valued vector representation $E_k^{w}$ for given a word $w_k$ can be obtained as follows:
\begin{equation}
E_k=M \cdot j(w_k)
\end{equation}
where $j(w_k)$ is the one hot vector representation of the word $w_k$. Thereafter, we augment the PoS and position embeddings (obtained from the previous layer) to the vector representation. 
\begin{equation}
x_k=E_k^{w} \oplus E_k^{PoS} \oplus E_k^{position}
\end{equation}
Where $E_k^{PoS}$ and $E_k^{position}$ are the PoS embedding and position embedding, respectively. The $\oplus$ denotes the concatenation operator. 
In our work, we use publicly available word embedding\footnote{http://bio.nlplab.org/} ($200$ dimensions) pre-trained on a combination of PubMed and PMC articles to the text extracted from a recent English Wikipedia dump. The performance of the word embedding depends on various hyperparameter setting such as vector dimension, context window size, learning rate, sample size etc. Pysallo et al. \cite{moen2013distributional} has released this pre-trained biomedical embedding after the deep analysis of various hyperparameter setting that obtains optimal embedding. Utilizing the pretrained word embedding not only helps in minimizing the time cost but also helpful in obtaining the best optimal parameter.
\subsection{Bi-LSTM unit}
Bi-directional LSTM consists of three layers as discussed below:
\subsubsection{Sequence Layer}\label{sequence-layer}
The sequence layer takes the input from embedding layer and provides a linear sequence as output. Recurrent neural network (RNN) is a powerful technique to encode a sentence by capturing long term dependency. However, because of the long sequence it often suffers with vanishing or exploding gradient problems \cite{rnn1,Pascanu2012UnderstandingTE}. This problem can be overcome by gating and memory mechanism as introduced in LSTM \cite{hochreiter1997long}. LSTM provides a different way to compute the hidden states. \\
\indent
The feature word sequence is represented by a bidirectional LSTM-RNNs \cite{rnn1}. The LSTM unit at $k$-th word consists of an \textit{input} gate $i_k$, \textit{forget} gate $f_k$, an \textit{output} gate $o_k$, a memory cell $c_k$ and hidden state $h_k$. The input to this unit is a $n$-dimensional input vector $x_k$, the previous
hidden state $h_{k−1}$, and the memory cell $h_{k−1}$, and computes the new hidden states as follows:
\begin{equation}
\begin{split}
i_{k} &=\sigma (W_{1}^{(i)}x_{k}+W_{2}^{(i)}h_{k-1} + b^{(i)} )  \\
f_{k} &=\sigma (W_{1}^{(f)}x_{k}+W_{2}^{(f)}h_{k-1} + b^{(f)} )  \\
o_{k} &=\sigma (W_{1}^{(o)}x_{k}+W_{2}^{(o)}h_{k-1} + b^{(o)} )  \\
u_{k} &=tanh(W_{1}^{(u)}x_{k}+W_{2}^{(u)}h_{k-1} + b^{(u)} )  \\
c_{k} &= i_{k} \odot u_{k}  +f_{k} \odot c_{k-1} \\
h_{k} &= o_{k} \odot tanh(c_{k})
\end{split}
\end{equation}
where $\sigma$, $ \odot$ denote the sigmoid function and element-wise multiplication, respectively. The $W_1$, $W_2$ and $b$ are the weight-matrix and bias vector, respectively. The LSTM unit
at $k-$th word feature takes the input as the concatenation of word embedding, PoS embedding and position embeddings obtained from an auto-encoder: $x_k=[E_k^{w} \oplus E_k^{PoS} \oplus E_k^{position}]$. We calculate the forward hidden state $\overrightarrow{h_k}$ and backward hidden state $\overleftarrow{h_k}$.
The final hidden state computed by augmenting both the hidden state $z_k=[\overrightarrow{h_k} \oplus \overleftarrow{h_k}]$. 

\subsubsection{Max-pooling Layer}
The max-pooling layer takes into account the hidden states of all the words in the SDP, instead of only the last word's hidden state. Max pooling takes the maximum over a position of the entire SDP sentence, hidden state obtained by concatenating the forward hidden state $\overrightarrow{h}$ and backward hidden state $\overleftarrow{h}$. Let us assume that a SDP sentence of length $N$ has a sequence of hidden states, $\{z_1,z_2 \ldots z_N\}$, then max-pooling layer computes the pooled hidden state $z_{max}^{i}$ at position $i$ as follows :
\begin{equation}
z_{max}^{i}=max(z_1^{i},z_2^{i} \ldots z_N^{i})
\end{equation}
Finally, the pooled hidden state $S$ for a SDP sentence is calculated by concatenating each of the pooled hidden state position values. The $S$ will be $[z_{max}^{1} \oplus \ldots \oplus z_{max}^{len} ]$, where $len$ denotes the dimension of the hidden state.
\subsubsection{Multilayer Perceptron (MLP) Model} \label{mlp}
The output of sequence layer $S$ is fed into a fully connected layer with $H$ number of hidden layers. More formally, given a sequence layer output $S$, number of hidden layers $H$, network calculates output as follows:

\begin{equation}
M=f(W_M*S+b_M)
\end{equation}
where $W_M \in \mathbb{R}^{H \times S}$ is the weight matrix between the output of sequence layer and hidden layer; $b_M \in \mathbb{R}^{H \times 1}$ is a bias term vector. \\
Thereafter, the output $M$ is transformed into $ T \in  \mathbb{R}^{L\times 1}$ by augmenting with a weight matrix $W_T \in \mathbb{R}^{L \times H}$, where $L$ is the number of required labels. In our case the value of $L=2$.
\begin{equation}
T=W_T* M
\end{equation}

Finally, the transformed output $T$ is fed into the softmax layer. The softmax layer provides the output probability of each label. Mathematically, it can be written as follows:  
\begin{equation} 
\begin{split}
P(L=l|T) & = \frac{e^{T_l}}{\sum_{k=1}^{L}e^{T_k}}
\end{split}
\end{equation}
\section{Results}\label{experiment}
\subsection{Dataset} \label{dataset}
The proposed model is evaluated on the two popular benchmark corpora for PPI, namely 
AiMed and BioInfer\footnote{http://corpora.informatik.hu-berlin.de/}. AiMed dataset is generated from $197$ abstracts extracted from the Database of Interacting Protein (DIP). It contains 
$1955$ sentences with the protein entities, manually tagged with the PPI interaction relations. This is recognized as the standard dataset for PPI extraction task.\\ 
\indent
The BioInfer corpus created by the Turku BioNLP group\footnote{http://bionlp.utu.fi/} consists of $836$ sentences. In our work, we assume the protein interacted pair as the positive instance and non-interacted pair as the negative instance. To identify the negative instances which are not directly given in the dataset, we assume all the possible pairs of proteins that are possible in a given sentence and consider those protein-pairs to be negative instances whose relations are not given in the sentence. Thereby, we obtain $3109$ negative instances and $939$ positive instances for AiMed corpus. Similarly, in case of BioInfer corpus, we obtain $5951$ negative instances over $1077$ positive interactions. It can be observed that both the datasets are imbalanced as they are strongly biased towards the negative examples. 
Statistics of these datasets are shown in Table-\ref{datasetStatistics}.
\subsection{Preprocessing}\label{preprcessing}
The protein entities are generalized with the protein IDs to make the model insensitive towards biases associated with the names of the proteins. This makes every protein unique and avoids the model to learn highly interacting protein pairs. We perform tokenization with the help of Genia Tagger\footnote{http://www.nactem.ac.uk/GENIA/tagger/}. The tokenized sentence is parsed with the Enju parser to obtain the dependency relations.
\begin{table}[h!]
\centering
{
\begin{tabular}{cccc}
 \hline
Dataset & Interacted Pair &  Non-interacted Pair & Ratio\\
 \hline
AiMed  & 939 & 3109 & 1:3.3\\
BioInfer & 1077 & 5951 & 1:5.5\\
 \hline
\end{tabular}%
}
\caption{Dataset Statistics for PPI Extraction}
\label{datasetStatistics}
\end{table}

\subsection{Network Training and Hyper-parameters}\label{training-objective}
The objective of training the LSTM model is to minimize the binary cross entropy cost function. It can be written as follows:
\begin{equation} \label{cost-function}
\small
\mathcal{L}(S, L) = -\frac{1}{n} \sum_{i=1}^n l^{(i)} \ln a(s^{(i)})+
\left(1 - l^{(i)}\right) \ln \left(1 - a(s^{(i)})\right)
\end{equation}
Here, $S=\{s^{(1)}, s^{(2)} \ldots s^{(n)}\}$ is the set of input SDP sentence in the training dataset, and  $L=\{l^{(1)}, l^{(2)} \ldots l^{(n)}\}$ is the corresponding set of labels for those SDP sentences. The $a(s)$  denote the output of the MLP layer. The gradient-based optimizer is used to minimize our cost function described in Eq-\ref{cost-function}. We have used Adam \cite{adam}, an adaptive learning rate based optimizer, to update the parameters throughout  training. To avoid over-fitting, the network dropout \cite{srivastava2014dropout} mechanisms are used with a dropout rate of $0.3$. \\
\indent
The hyper-parameter values were determined from the preliminary experiments by evaluating the model performance for $10$-fold cross-validation. 
The proposed model described in Section-\ref{architecture} is implemented using Keras\footnote{https://keras.io/}. We have chosen Tensorflow as backend machine learning library.
We tune our model for various hyper-parameters of the LSTM architecture including the number of LSTM units, dropout ratio, number of epochs and different optimization algorithms etc. for both the datasets. 
We obtain the best results for both the AiMed and BioInfer datasets on a set of optimized network hyper-parameters. 
This reflects the generalization of our optimum hyper-parameter selection over two completely different datasets. Table \ref{parameter-value} provides the details about the optimal hyperparameter settings using 10-fold cross validation experiments.


\begin{table}[]
\centering
\begin{tabular}{|l|l|l|}
\hline
\textbf{Activation Function} & \textbf{F-Score (AiMed)} & \textbf{F-Score (BioInfer)} \\ \hline
Sigmoid & 86.45 & 77.35 \\ \hline
ReLU & 85.77 & 76.92 \\ \hline
tanh & 84.40 & 75.79 \\ \hline
\end{tabular}
\caption{Comparison of different activation functions with same hyperparameters values}
\label{activation}
\end{table}

\begin{minipage}{0.98 \textwidth}
  \begin{minipage}[b]{0.45\textwidth}
    \begin{tabular}{cc} 
 \hline
 \textbf{Hyper-parameters} & \begin{tabular}[c]{@{}c@{}} \textbf{Optimal}\\ \textbf{value} \end{tabular}\\ \hline
 Number of LSTM units & $64$ \\ 
 Dropout ratio & $0.3$  \\
 Activation function & Sigmoid \\
 Optimization algorithm & Adam\\
\# Epochs & $130$  \\
Size of MLP layer output &  $30$   \\
 \hline
\end{tabular}
    \captionof{table}{Optimal hyper-parameter setting on $10$-fold cross validation for both the datasets.}
\label{parameter-value}
  \end{minipage}
  \hfill
  \begin{minipage}[b]{0.5\textwidth}
    \centering
    \resizebox{\textwidth}{!}{%
    \begin{tabular}{|c|c|c|}
\hline
\begin{tabular}[c]{@{}c@{}} \textbf{Context window}\\ \textbf{size} \end{tabular}  & \begin{tabular}[c]{@{}c@{}} \textbf{F-score}\\ \textbf{(AiMed)} \end{tabular} & \begin{tabular}[c]{@{}c@{}} \textbf{F-score}\\ \textbf{(BioInfer)} \end{tabular} \\ \hline
{[}-5,5{]} & 78.36 & 72.72 \\ \hline
{[}-6,6{]} & 78.54 & 73.18 \\ \hline
{[}-7,7{]} & 79.19 & 73.23 \\ \hline
{[}-8,8{]} & 81.16 & 74.29 \\ \hline
{[}-9,9{]} & 81.75 & 75.56 \\ \hline
{[}-10,10{]} & 82.89 & 75.93 \\ \hline
{[}-11,11{]} & 82.17 & 75.28 \\ \hline
{[}-12,12{]} & 81.41 & 74.88 \\ \hline
\end{tabular}
}
      \captionof{table}{Analysis of context window on 10 fold cross validation data for position feature}
\label{context-window}
    \end{minipage}
  \end{minipage}


\subsection{Analysis of Hyper-parameter Settings}
We setup all the experiments by varying the hyper-parameter values and analyze the behaviors of our model. For AiMed dataset, we observed that addition of LSTM units improves the model performance to a certain extent. Thereafter, it keeps on decreasing gradually. We define an optimal value $64$ for the same, via cross-validation experiment. It was observed that
deep MLP layer helps to improve the overall performance of the model when compared to a shallow MLP layer as shown in Figure 3. However, this improvement was dependent upon the size of the output layer, which increases initially from $20$ to $30$ and then starts decreasing while the size of the output layer is increased to $50$. We also notice that stacking of another MLP layer makes our model complex, thus reducing the performance on a cross-validation setting.\\
\indent In case of BioInfer dataset, we also observe quite a similar trend in performance with the addition of LSTM units, size of MLP output layer and stacking of another MLP layer. The optimal values are reported in Table-\ref{parameter-value}.
We also analyze the performance of our model on the number of epochs for which training was performed on both the datasets. On AIMed dataset, the value of F1-score initially shows a short fall with the increasing number of epochs from $1$ to $2$ and then shows regular growth with the increasing number of epochs from $2$ to $130$, and finally a dip on further increasing the number of epochs to $135$ and $140$. This can be attributed to the fact that training on a very large number of epochs makes our model over-fitted and hence the cross-validation accuracy decreases. There has also been an initial decline in the F1-score in case of BioInfer dataset but then there has been steady increase with the increase in the number of epochs. We achieve the optimum performance with the same number of epochs ($130$) for both the datasets.
To further compare the performance of ReLU with Sigmoid, we have also conducted the experiments considering both ReLU and Sigmoid on both the datasets. On both the datasets, the sigmoid function was found to be superior over ReLU. The results are reported in Table-\ref{activation}.
\begin{minipage}{\textwidth}
  \begin{minipage}[b]{0.45\textwidth}
    \centering
    \includegraphics[height=5cm]{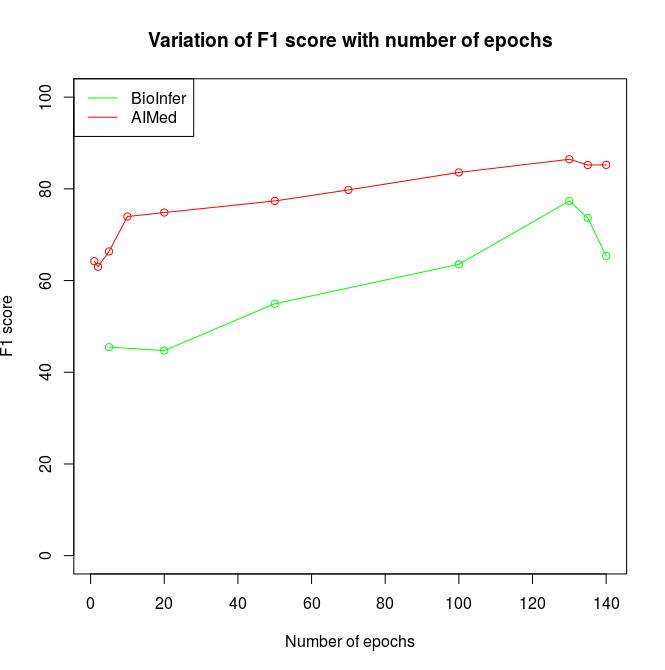}
    \captionof{figure}{Effect of varying MLP layers and their output sizes}
    \label{MLP}
  \end{minipage}
 \hfill
  \begin{minipage}[b]{0.45\textwidth}
    \centering
    \includegraphics[ height=5cm]{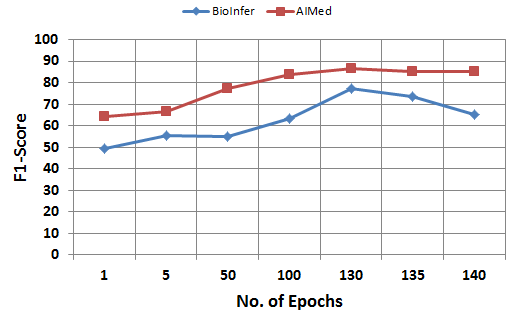}
      \captionof{figure}{Effect of varying epochs on F1-Score}
      \label{epoch}
    \end{minipage}
  \end{minipage}


\subsection{Evaluation on Benchmark Datasets}
In the recent years, different kernel-based techniques and SVM based model were adopted as baselines against the deep learning CNN based model for the PPI task. It has been shown how deep learning based models perform superior compared to the feature based models 
\cite{kim:2014:EMNLP2014,choi2016extraction}. As such, in order to make an effective comparison of our proposed approach, we design two strong baselines based on neural network architecture as follows:\\
\textbf{Baseline 1:} The first baseline model is constructed by training a multi-layer perceptron on the features obtained from the embedding layer as defined in subsection-\ref{embedding-layer}. The sentence embedding $S_M$ is generated by the concatenation of every PoS and position augmented word embeddings to SDP embedding.
\begin{equation}
S_M=x_1 \oplus x_2 \ldots \oplus x_n;
\end{equation}
Thereafter, $S_M$ is fed into MLP layer described in Subsection-\ref{mlp}.\\
\textbf{Baseline 2:} Our second baseline is based on the more advanced sentence encoding techniques, RNN. The SDP sentence encoding $S_R$ can be generated as follows: 
\begin{equation}
S_R= \sigma(\mathbf{U}*x_n + \mathbf{V}*h(n-1) + \mathbf{b})
\end{equation}
where $\sigma$ is a sigmoid function, $h(n-1)$ denotes the hidden representation of $(n-1)^{th}$ word in the SDP sentence. \textbf{$U$}, \textbf{$V$}, and \textbf{$b$} are the network parameters. Similar to Baseline 1, MLP layer is used to classify a SDP sentence into one of the two classes, \textit{viz}: `interacting pair' and `non-interacting pair'. \\

\indent We perform 10-fold cross validation on both the datasets. With no official development data set available, cross validation seems to be the most reliable method of evaluating our proposed model. To evaluate the performance of our model, we use standard recall, precision, and F1-score. The detailed comparative analysis of our proposed model (sdpLSTM) over these baselines and state-of-art systems are reported in Table-\ref{results}. The obtained results clearly show the effectiveness of our proposed sdpLSTM based model over the other models exploring neural network architectures or conventional kernel or SDP based machine learning model. The optimum performance is achieved with $130$ epochs for both the datasets as depicted in Figure-\ref{epoch}. Statistical significance tests verify that improvements over both the baselines are statistically significant as (\textit{p-value} $<$ $0.05$). In our proposed model we obtain the significant F1-score improvement of $19.99$ and $16.23$ points over the first two baselines for the AiMed dataset, respectively. On BioInfer dataset, our system shows the F1-Score improvement of $7.13$ and $4.14$ points over these two baselines, respectively. 
\section{Analysis}
\subsection{Comparative Analysis with Existing Methods}
In order to perform the comparative analysis with the existing approaches, we choose the recent approach exploiting neural network model. We explore other approaches utilizing SVM based kernel methods and word embedding feature as shown in Table-\ref{results}. 
We observe that sdpLSTM significantly performs better than all the state-of the-art techniques for both the datasets. From this, we can conclude that sdpLSTM is more powerful in extracting protein interacted pairs over the CNN based architecture developed in \cite{hua2016shortest} and \cite{choi2016extraction}. We further make an interesting observation that incorporating the latent features embedded into the neural network based architecture improves the performance of the system.

Our proposed model attains an improvement of $1.25$ F-score point (c.f. Table 8) over the model proposed in \cite{choi2016extraction} for the AiMed dataset.  However, it should be noted that DCNN model made use of a significant number (total $29$) of domain dependent lexical, syntax and semantic level features. 
In contrast to this our model is more generic in the sense that we use only PoS and position features. We further re-implemented the DCNN system and evaluated it on both the datasets. Evaluation (c.f. Table 8) shows that our proposed model attains better performance for both the datasets. 
We also re-implemented the system reported in \cite{li2015approach} 
to obtain the precision and recall values. The observed improvements over the existing systems are statistically significant (as \textit{p-value} $<$ $0.05$).

\begin{table*}[]
\centering
\resizebox{\textwidth}{!}{%
\begin{tabular}{llcccccc}
\hline
\multirow{2}{*}{\textbf{Model}} & \multicolumn{1}{c}{\multirow{2}{*}{\textbf{Approach}}} & \multicolumn{3}{c}{\textbf{AIMED}} & \multicolumn{3}{c}{\textbf{BioInfer}} \\ \cline{3-8} 
 & \multicolumn{1}{c}{} & \textbf{Precision} & \textbf{Recall} & \textbf{F1-Score} & \textbf{Precision} & \textbf{Recall} & \textbf{F1-Score} \\ \hline
Baseline 1 & MLP (SDP+Feature Embedding) & 59.73 & 75.93 & 66.46 & 68.56 & 72.05 & 70.22 \\ 
Baseline 2 & RNN (SDP+Feature Embedding) & 66.23 & 74.72 & 70.22 & 71.89  & 74.59 & 73.21 \\ \hline
\rowcolor{Gray}
Proposed Model & sdpLSTM (SDP+Feature Embedding) & 91.10 & 82.2 & 86.45 & 72.40 & 83.10 & 77.35 \\ \hline
\cite{hua2016shortest} & sdpCNN (SDP+CNN) & 64.80 & 67.80 & 66.00 & 73.40 & 77.00 & 75.20 \\ 
\cite{choi2016extraction} & \begin{tabular}[c]{@{}l@{}}DCNN (CNN+word/position embeddings+\\ Semantic (WordNet) feature embeddings)\end{tabular} & - & - & 85.2 & - & - & - \\ 
\cite{choi2016extraction}$^{*}$ & \begin{tabular}[c]{@{}l@{}}DCNN \end{tabular} & 88.61 & 81.72 & 85.03 & 72.05 & 77.51 & 74.68 \\ 
\cite{qian2012tree} & Single kernel+ Multiple Parser+SVM & 59.10 & 57.60 & 58.10 & 63.61 & 61.24 & 62.40 \\ 
\cite{peng2017deep} & \begin{tabular}[c]{@{}l@{}}McDepCNN (CNN+word+PoS+Chunk+NEs\\ Multi-channel embedding)\end{tabular} &  67.3 & 60.1 & 63.5 & 62.7 &  68.2 & 65.3 \\
\cite{zhao2016protein} & \begin{tabular}[c]{@{}l@{}}Deep neutral network\end{tabular} &  51.5 &  63.4 & 56.1 & 53.9 &  72.9 & 61.6 \\
\cite{tikk2010comprehensive} & \begin{tabular}[c]{@{}l@{}}All-path graph kernel\end{tabular} &   49.2 &  64.6 & 55.3 & 53.3 & 70.1 &  60.0 \\
\cite{li2015approach} & Multiple kernel+ Word Embedding+ SVM & - & - & 69.70 & - & - & 74.0 \\
\cite{li2015approach}$^{*}$ & Multiple kernel+ Word Embedding+ SVM & 67.18 & 69.35 & 68.25 & 72.33 & 74.94 & 73.61 \\
\cite{choi2010simplicity} & Tuned tree kernels +SVM & 72.80 & 62.10 & 67.00 & 74.5 & 70.9 & 72.60 \\ \hline
\end{tabular}%
}
\caption{Comparative results of the proposed model (sdpLSTM) with different baselines and state-of-the-art systems.  Ref. \cite{choi2016extraction}$^{*}$ and \cite{li2015approach}$^{*}$ denote the reimplementation of the systems proposed in \cite{choi2016extraction} and \cite{li2015approach} with the authors reported experimental setups.}
\label{results}
\end{table*}
\vspace{-1em}
\subsection{Effects of Feature Combination}
In this section, we analyze the significance of each feature by performing feature augmentation (including feature one by one) as shown in Table-\ref{feature-combination}. We begin by examining only SDP embedding. It can be observed that SDP based embedding alone shows a remarkable performance of $86.38$ and $75.86$ F1-Score on AiMed and BioInfer dataset, respectively. 
This clearly shows the significance of SDP based embedding in identifying protein interacted pairs. However, we observe that inclusion of PoS and position embedding does not have any positive impact on the AiMed dataset.
In fact, there have been drops in F1-score by $0.12$ point when PoS feature is added and $3.49$ point when position embedding feature is included. This might be due to the data sparseness problem with the lack of training data. In case of BioInfer dataset, PoS tagging is comparatively less informative, but still boosts the F1-score by $0.54$ F1-score points. The inclusion of position embedding, however, shows very minor improvement of $0.07$ F-score points.
The reason is while adding a position to PoS feature, it helps as we have PoS tag information (which is NNP) of the closest potential entity. 
We analyze that, the improvements are not simply due to combining the features to SDP embedding. This suggests that these information sources are complementary to each other in some linguistic aspects. We closely investigate the outputs of AiMed dataset produced in our system and summaries with the following observations:\\
\begin{enumerate}
\item \textbf{PoS distribution:} Protein names are mainly noun phrases. For the AiMed dataset, we observed that the multi-word proteins were not properly tagged as the noun phrases. This has encountered some errors which propagated when we have introduced the PoS alone as a feature to the LSTM model.
\item \textbf{Presence of protein interacted words:} The presence of protein interacted words (inhibit, regulated, interaction etc.) provides an important clue to identify the interaction of proteins. When the system takes SDP as input, we observe that in some cases the PoS tagger is unable to tag the interacted words as verbs. We did quantitative analysis and found that the PoS tagger could not tag verb phrases correctly in $316$ SDP sentences out of total $4048$ sentences. This could be one of the reasons that system performance is comparable when we use the PoS information alone as a feature.
\item \textbf{Position feature:} The position feature is a helpful feature to capture the most important words occurring in the vicinity of the protein words. 
We observe that length of input SDP sentences for incorrectly classified instances were higher compared to the correctly classified SDP sentence. Another observation of misclassified instances was, the existence of multiple protein entities in such cases. 
\end{enumerate}

We further investigate the reason for a decrease in the recall value on both the dataset by the incorporation of position embedding to SDP as shown in Table-\ref{feature-combination}. Our analysis revealed that position embedding feature was unable to capture the implicit form of PPI information that occurs in the vicinity of the window size. For example, \textit{``In cells lacking the \textbf{myosin-II heavy chain}, the bundles, which were induced by an over-expression of \textbf{cofilin}, shortened and became straight following hyperosmotic stress, forming a polygonal structure.''} In this example, there is no explicit mention of the protein interacted verb information, like `\textit{bind}', `\textit{interact}' etc. which makes difficult to capture the relevant words via position embeddings.
We identified that a total of $57$, $29$ cases where our system fails to capture the implicit form of protein information and incorrectly predicting it as non-interacting pair on BioInfer and AiMed dataset respectively.
Interestingly, combination of all the features modestly improves the performance of the system by $1.49$ and $0.07$ F-score points on BioInfer and AiMed dataset, respectively. 
We observe that when the model is evaluated on the less epoch, performance improvement with the addition of features is $1\%$-$2\%$. While increasing the epoch vanishes the impact of additional features.

\begin{table}[]
\centering
\resizebox{\textwidth}{!}{%
\begin{tabular}{ccccccc}
\hline
\multicolumn{1}{c}{\multirow{2}{*}{\textbf{Model}}} & \multicolumn{3}{c}{\textbf{AiMed}} & \multicolumn{3}{c}{\textbf{BioInfer}} \\ \cline{2-7} 
\multicolumn{1}{c}{} & \textbf{Precision} & \textbf{Recall} & \textbf{F1-Score} & \textbf{Precision} & \textbf{Recall} & \textbf{F1-Score} \\ \hline
Bi-LSTM(SDP) & 82.25 & 91.11 & 86.38 & 71.51 & 81.06 & 75.86 \\ 
Bi-LSTM(SDP+PoS Embedding) & 81.41 & 91.91 & 86.26 & 71.50 & 82.08 & 76.40 \\ 
Bi-LSTM(SDP+Position Embedding) & 77.56 & 90.86 & 82.89 & 72.08 & 80.60 & 75.93 \\ \hline
\rowcolor{Gray}
Bi-LSTM(SDP+PoS+Position Embedding) & 91.10 & 82.20 & 86.45 & 72.40 & 83.10 & 77.35 \\ \hline
\end{tabular}%
}
\caption{Analysis of feature combination}
\label{feature-combination}
\end{table}
\vspace{-1em}
\subsection{Error Analysis}
In this subsection, we analyze different sources of errors which lead to misclassification. We closely study the false positive and false negative instances and come up with following observations: \\
\textbf{(1)} When Enju dependency parser fails to capture dependencies, the error is propagated to BFS algorithm as such it does not return any valid SDP. For example, in the given sentence\\
``\textit{The ProtId1 or ProtId2 family is targets of cytokines and other agents that induce HIV-1 gene expression}", the SDP output mentioned are \textbf{``ProtId1 and ProtId2"} and \textbf{``ProtId1 family ProtId2"}. It should be noted that this is a negative example and our SDP fails to capture the context. This hampers our accuracy significantly.\\
\textbf{(2)} Interaction bearing words carry important information to identify protein interacted pairs such as bind, interact, inhibit. However, when interaction bearing words appear in negative context, system fails to properly classify it as non-interacted protein pair. For e.g. \textit{``in GSK-3 inhibitors suppressed Sema4D-induced growth''}, \textbf{inhibit} does not occur here in context of PPI.
\section{Conclusion and Future Works}
In this article, we have proposed an efficient model based on deep learning technique for PPI. The model makes use of SDP embeddings as feature. In addition it also exploits the latent PoS and position embedding features to complement the SDP embedding.
The main contribution of the proposed
methodology is the systematic integration of word embeddings learned from the biomedical literature and the use of SDP between protein pairs into the deep sdpLSTM architecture. 
Bio-medical word embedding was observed to capture semantic information
more effectively than internal embedding. By employing SDP
and LSTM, the proposed approach could make full use of structural information.
Our comprehensive experimental results on two benchmark biomedical corpora, AiMed and BioInfer demonstrated that (i) the SDP based word embedding input is effective to describe protein-protein relationship in PPI
task; (ii) the LSTM architecture is useful to capture the long contextual 
and structure information; and (iii) high-quality pretrained word
embedding is important in PPI task. The obtained results depict the superiority of sdpLSTM over the complex state-of-art approaches leveraging CNN and several higher level features with the significant F1-score improvements of $1.25$ and $2.15$ points on AiMed and BioInfer dataset, respectively.\\
In future, we would like to validate our approach on other relation extraction tasks such as drug-drug interaction, chemical-protein interaction by overcoming the possible errors. Multi-layer Bi-LSTM model has shown tremendous success in machine translation. One interesting direction of future work will be to develop a multi-layer Bi-LSTM model for relation extraction. Further, owing to the capability of attention mechanism, we would experiment by attentive pooling. \\

\begin{large}
\textbf{References}
\end{large}\\
\bibliographystyle{elsarticle-num}
\bibliography{PPI}

\end{document}